\documentstyle[11pt,newpasp,twoside,epsf]{article}
\begin{document}
\title{Properties of galaxy disks in hierarchical hydrodynamical simulations:
  Comparison with observational data}
\author{A. S\'aiz, R. Dom\'{\i}nguez-Tenreiro}
\affil{Depto.\ de F\'{\i}sica Te\'orica, C-XI.\@ Universidad Aut\'onoma de
  Madrid, Madrid, E-28049, Spain}
\author{P. B. Tissera}
\affil{I.A.F.E., Casilla de Correos 67, Suc.\ 28, Buenos Aires, 1428,
  Argentina}
\author{S. Courteau}
\affil{University of British Columbia, Dept.\ of Physics and Astronomy,
  Vancouver, BC, Canada V6T 1Z1}
\begin{abstract}
  We analyze the structural and dynamical properties of disk-like objects
  formed in fully consistent cosmological simulations with an
  inefficient star
  formation algorithm. Comparison with data of similar observable properties
  of spiral galaxies gives satisfactory agreement.
\end{abstract}
We present results of a detailed comparison
between the parameters characterizing the structural and dynamical
properties of a sample of 29 simulated disk-like objects (DLOs)
and those measured
in observed spiral galaxies. The DLOs have been identified in AP3M-SPH
fully consistent hierarchical hydrodynamical simulations,
where an {\em inefficient\/} Schmidt law-like algorithm to model the
stellar formation processes has been implemented.

A surface density bulge-disk decomposition was performed
on the DLOs, using a double-exponential profile. The
resulting
bulge and disk scale lengths,
$R_{\rm b}$ and $R_{\rm d}$,
and their ratio $R_{\rm b}/R_{\rm d}$, are consistent with available data
(Courteau, de Jong, \& Broeils 1996; de Jong 1996;
Moriondo, Giovanelli, \& Haynes 1999). See Fig.~1(a).
\begin{figure}
  \plottwo{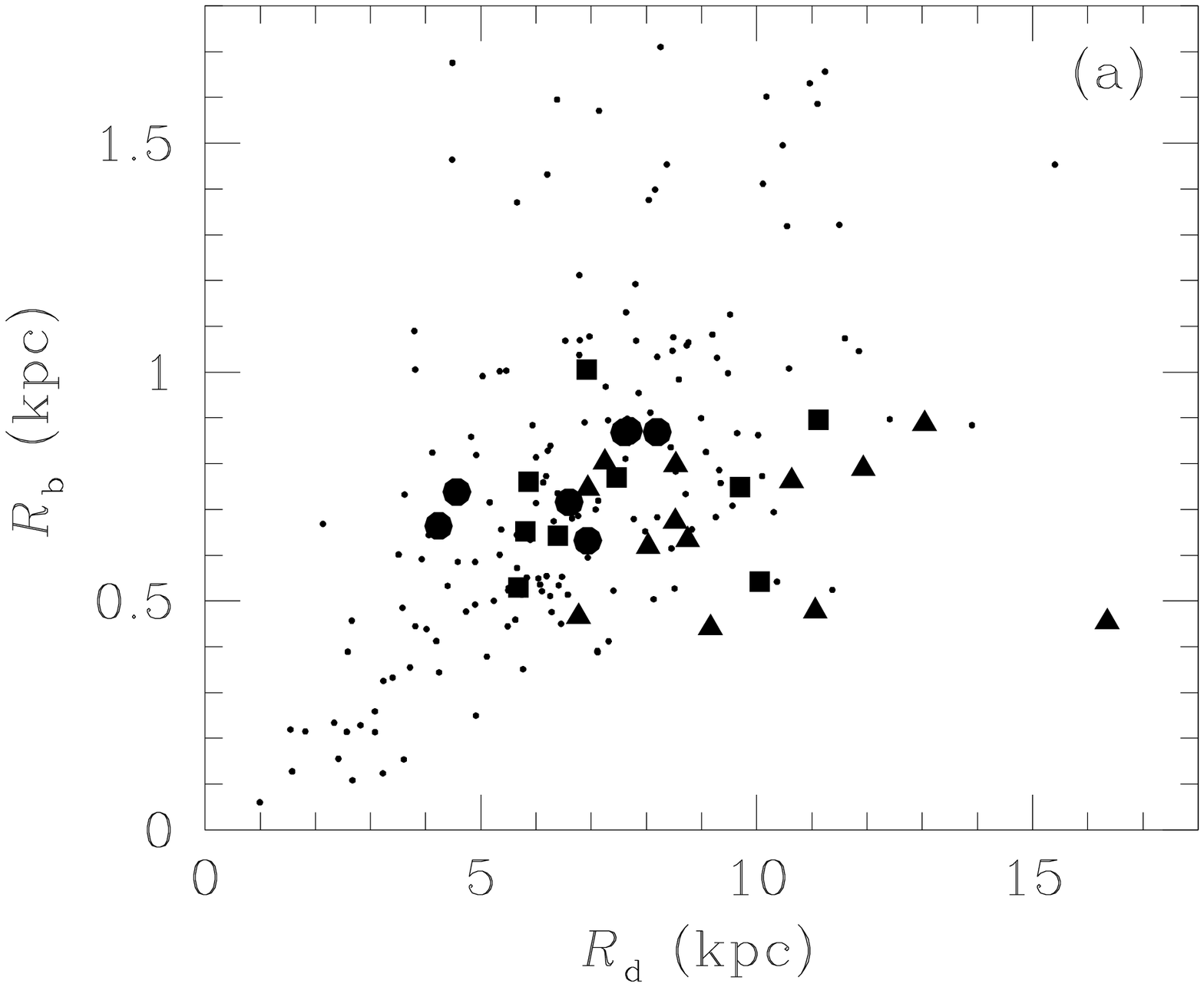}{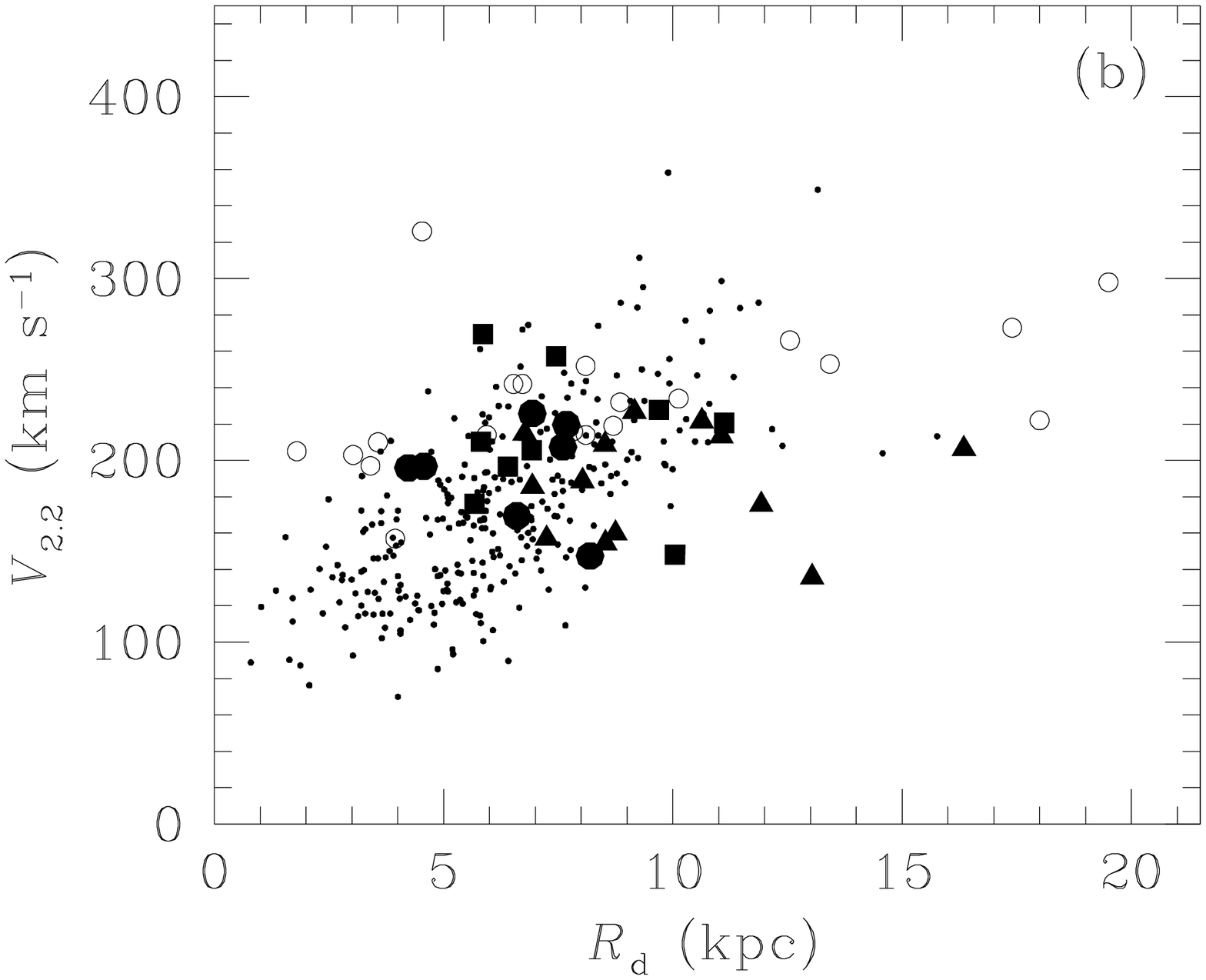}
  \caption{The scale lengths $R_{\rm b}$ and $R_{\rm d}$ (a),
    and the $V_{2.2}$ velocities versus $R_{\rm d}$ (b).
    Filled symbols correspond to
    DLOs formed in our simulations. Dots in (a) are from 1D decompositions
    of surface brightness profiles given by Courteau, de Jong, \& Broeils
    (1996) and Courteau (1997). Dots in (b) are data from the Courteau--Faber
    (H$\alpha$ spectra) sample. Open circles are data from spirals with
    \hbox{H\,{\sc i}} rotation curves from Casertano \& van Gorkom (1991).}
  \vspace{-0.3cm}
\end{figure}

Our rotation curve 
analysis and shape modeling uses a
parameterization based on the spatial scale
$R_{2.2} = 2.2 \, R_{\rm d}$ 
and the velocity (mass) estimate 
$V_{2.2} = V_{\rm cir} (R_{2.2})$. 
These parameters were compared with those measured for the
Courteau--Faber sample of bright Sb--Sc field spirals with
long-slit H$\alpha$ rotation curves, 
and the Broeils
(1992) compilation of late-type spirals with extended 
\hbox{H\,{\sc i}} 
rotation curves.  
Deep surface photometry is available for both samples.
In contrast
to findings
in other fully consistent
hydrodynamical simulations
(Navarro \& Steinmetz 2000 and references therein),
we find DLO $V_{2.2}$
values that are consistent with the observational data.
This is a consequence of disk formation
with conservation of the specific angular momentum $j$.
See Fig.~1(b).

In conclusion, the comparison between DLOs produced in our simulations and
observational data allows us to affirm that 
DLOs
have counterparts
in today spiral galaxies.
This agreement suggests that the process
operating in Fall \& Efstathiou (1980) standard
model for disk formation (i.e., gas cooling and collapse with
$j$ conservation) is also at work in the
{\em quiescent\/} phases of DLO formation in these simulations. 
However,
{\em violent\/} episodes (i.e., interactions and merger events) also occur
and play an important role in DLO assembly (Dom\'{\i}nguez-Tenreiro,
Tissera, \& S\'aiz 1998). 
To provide the right conditions for disk regeneration
after the last violent episode of DLO assembly, a compact central stellar
bulge is needed; this will ensure the axisymmetric character
of the gravitational potential well at scales of some kpcs at all
times, avoiding excessive $j$ losses in violent events.
A second condition is necessary: the availability of gas at low~$z$,
to form the disk.
The good match of our DLO parameters with obervational
data suggests that our {\em inefficient \/} star formation
algorithm meets both requirements.
This global agreement with observations also represents
an important step towards making numerical approaches more widely used
for the study of
galaxy formation and evolution
in a cosmological framework, i.e., from primordial fluctuations.

\end{document}